\documentclass[a4paper]{jpconf}
\usepackage{graphicx}
\usepackage{amsmath, amsthm, graphicx,amssymb,subfigure,bbm}
\begin{document}
\title{Quantised relativistic membranes and non-perturbative checks of
gauge/gravity duality}

\author{Veselin G. Filev ${}^1$ and Denjoe O'Connor${}^2$}
\address{${}^1$  Institute of Mathematics and Informatics,\\
Bulgarian Academy of Sciences, Acad. G. Bonchev Str., 1113 Sofia, Bulgaria}
\address{${}^2$ School of Theoretical Physics,
  Dublin Institute for Advanced Studies, \\
       10 Burlington Road, 
       Dublin 4, Ireland.
}

\ead{denjoe@stp.dias.ie, vfilev@math.bas.bg}

\begin{abstract}
  We test the background geometry of the BFSS model using a D4--brane
  probe. This proves a sensitive test of the geometry and we find
  excellent agreement with the D4--brane predictions based on
  the solution of a membrane corresponding to the D4--brane propagating on
  this background. 
\end{abstract}

\section{Introduction}
From the perspective of a quantised eleven dimensional supermembrane
it is perhaps natural to argue that there ought to be a dual
gravitational background to the matrix regulated supermembrane
proposed by Banks et al \cite{Banks:1996vh} as a non-perturbative
formulation of $M$-theory.  However, it is important to test this
hypothesis as far as possible.  In recent years several
non-perturbative tests of this hypothesis have been performed
\cite{Anagnostopoulos:2007fw,Catterall:2008yz,Hanada:2008ez,Hanada:2013rga,Kadoh:2015mka,Filev:2015hia,Hanada:2016zxj,Berkowitz:2016jlq}
at the level of comparing the expectation value of the microscopic
Hamiltonian with the energy prediction coming from the black hole
background. These are increasingly sensitive tests and the agreement
found between theory and simulation in the most recent, and currently
most precise test, by Berkowitz et al \cite{Berkowitz:2016jlq} while
impressive is open to question. They observe that if they fit their
data ``by a single powerlaw $E_0(T)=a T^p$, then $E_0(T)= (3.13\pm
0.03) T^{2.02\pm0.03} \simeq\pi T^2$ describes our continuum large-N
data very well ($\chi^2/DOF = 7.7/5$) in the whole temperature
range.'' While their precision tests are still rather convincing it is
important to have other tests that check the geometry in more detail.
This can be done, just as one does in practice in the everyday world,
by using probes which are sensitive to the geometry yet do not in turn
change it significantly.\footnote{A similar study using probe
  D0-branes has been performed in ref.~\cite{Rinaldi:2017mjl}. }This
corresponds to adding $M5$-membrane density to the BFSS matrix model
-- the Berkooz-Douglas (BD) matrix model. From a string theory point
of view the BD model describes the low energy regime of the
D0/D4--brane intersection. It is a flavoured $1+0$ dimensional
gauge theory with ${\cal N}=8$ supersymmetry and is the dimensional
reduction of $4-d$  ${\cal N}=2$ supersymmetric Yang--Mills theory.
This makes the BD model particularly interesting in the context of holographic
studies of flavour dynamics, since the ${\cal  N}=16$ supersymmetric
Yang--Mills theory which it probes has a well know gravitation dual
background.

The proposal that there is a gravitational dual to gauged matter
systems represents a dramatic insight into certain non-perturbative
phenomena. In the original formulation of Maldacena~\cite{Maldacena:1997re}, 
the duality relates string theory in the $AdS_5\times S^5$ background space-time 
to the large $N$ limit of $3+1$ dimensional ${\cal N}=4$ Supersymmetric Yang-Mill
theory living on the asymptotic boundary of the AdS$_5$
space-time. This idea has inspired numerous extensions of the duality
with ever increasing phenomenological relevance, currently ranging
from heavy ion collisions to condensed matter physics. In this paper
we are interested in holographic flavour dynamics--the generalisation
of the AdS/CFT correspondence to flavoured gauge theories.

The first such generalisation was proposed by Karch and
Katz~\cite{Karch:2002sh} , who introduced a probe D7--brane to the
$AdS_5\times S^5$ supergravity background. On the field theory side
this corresponds to introducing an ${\cal N}=2$ fundamental
hypermultiplet in the quenched approximation. The classical dynamics
of the probe brane is governed by an effective Dirac-Born-Infeld
action. Remarkably the AdS/CFT dictionary relates the classical
properties of the brane to quantum vacuum expectation values in the
dual flavoured gauge theory. One such quantity is the fundamental
condensate of the theory, which is encoded in the classical profile of
the probe brane near the asymptotic boundary. In
refs.~\cite{Babington:2003vm} and \cite{Hoyos:2006gb} the finite
temperature set-up has been considered. The authors uncovered a first
order meson melting phase transition corresponding to a topology
change transition of the possible D7--brane embeddings. In ref.~\cite{Mateos:2007vn} 
these studies have been extended to the general
Dp/Dq--brane system and certain universal properties of the
corresponding holographic gauge theories have been uncovered. In fact the BD matrix model is dual to
the D0/D4--brane system, which falls into the same universality class as the phenomenologically relevant D3/D7--brane
system.

Another attractive feature of the BD matrix model is that it is $1+0$
dimensional supersymmetric Yang--Mills theory, which makes it super
renormalizable and thus accessible with lattice
simulations. Therefore, the BD model can serve as a bridge between
lattice gauge theory and holography allowing a direct precision test
of the gauge/gravity duality with flavours. This was our main
motivation to study the D0/D4 system and the BD matrix model
\cite{Filev:2015cmz}.

Our studies provide a highly non-trivial test of the AdS/CFT
correspondence with matter.  The results provide substantial evidence
for the validity of the holographic approach to flavour dynamics.
They also test the predicted D0-brane geometry in a highly
non-trivial way since as the fundamental mass is varied, the D4-brane
probes the radial dependence of this dual geometry. The agreement
we find with the predictions from the D4-brane probe embedded in the
dual D0 black hole geometry is remarkable. Although it is not a
mathematical proof, we believe that the remarkable agreement between
theory and simulation, which we uncovered is due to the cancellation of
$\alpha'$ corrections in the black hole embedding.

\section{Holographic flavours in one dimension}

In this section we review the description of the D0/D4--brane system  in the quenched
approximation adapting the general discussion of references~\cite{Mateos:2007vn} and~\cite{Itzhaki:1998dd}.

\subsection{D0-brane background}
In the near horizon limit the D0-brane background is given by the
metric:
\begin{eqnarray}\label{metric}
ds^2&=&-H^{-\frac{1}{2}}\,f\,dt^2+H^{\frac{1}{2}}\left(\frac{du^2}{f}+u^2\,d\Omega_8^2\right)\ ,\nonumber \\
e^{\Phi}&=&H^{\frac{3}{4}}\ ,~~~~~~C_0 =H^{-1}\ ,
\end{eqnarray}
where $H=\left(L/u\right)^7$, $f(u) = 1-(u_0/u)^{7}$ is the blackening factor, $\Phi$ is the dilaton field and $C_0$ is the only component of the RR one form coupled to the D0-branes . Here $u_0$ is the radius of the horizon and the length scale $L$ can be expressed in terms of string theory units as:
\begin{equation}
L^7=60\,\pi^3\,g_s\,N_c\,\alpha'^{7/2}\ ,
\end{equation}
where $N_c$ is the number of D0--branes corresponding to the rank of the gauge group of the dual field theory\footnote{Note that we will abbreviate $N_c$ to $N$ when the context is clear.}. According to the general gauge/gravity duality \cite{Itzhaki:1998dd}, the Yang-Mills coupling of the corresponding dual gauge theory is given by:
\begin{equation}
g_{\rm YM}^2=g_s\,(2\pi)^{-2}\,\alpha'^{-3/2}\ .
\end{equation}
The Yang-Mills coupling is dimensionful and the corresponding dimensionless effective coupling runs with the energy scale according to:
\begin{equation}\label{geff}
g_{\rm eff}^2=\lambda\,U^{-3}\ ,
\end{equation}
where $\lambda = g_{\rm YM}^2\,N_c$ is the t'Hooft coupling. The supergravity background can be trusted if both the curvature and the dilaton are small, which leads to the restriction \cite{Itzhaki:1998dd}:
\begin{equation}\label{bounds}
1\ll g_{\rm eff} \ll N_c^{\frac{4}{7}}\ .
\end{equation}
and the theory is strongly coupled in this regime. From equations (\ref{metric}) and (\ref{geff}) it follows that the upper bound in equation (\ref{bounds}) can be violated at low energies (small radial distances) when the dilaton blows, however at finite temperature and fixed 'tHooft coupling, $g_{\rm eff}$ peaks at the black hole horizon and the bound $\lambda/T^3 \ll N_c^{8/7}$ is satisfied in the large $N$ limit. At high energies (large radial distances) the curvature of the background grows, while the effective coupling decreases. As a result the lower bound in (\ref{bounds}) is violated at energies higher than approximately $\lambda^{1/3}$ and hence $\alpha'$ corrections are increasingly important at large radial distances.

Finally, the Hawking temperature of the background is given by:
\begin{equation}
T =\frac{7}{4\,\pi\,L}\left(\frac{u_0}{L}\right)^{\frac{5}{2}}
\end{equation}
and is identified with the temperature of the dual gauge theory.

\subsection{Flavour D4-branes}
\label{holo-condensate}
To introduce matter in the fundamental representation we consider the addition of $N_f$ D4-branes to the D0-brane background. In the probe approximation $N_f\ll N_c$, the dynamics of the D4-branes is governed by the Dirac-Born-Infeld, which in the absence of a background B-field is given by:
\begin{equation}\label{DBI}
S_{\rm DBI} = -N_f\,T_4\int\,d^4\xi\,e^{-\Phi}\,\sqrt{-{\rm det}||G_{\alpha,\beta}+(2\pi\alpha')F_{\alpha,\beta}||}\ ,
\end{equation}
where $G_{\alpha,\beta}$ is the induced metric and $F_{\alpha,\beta}$ is the $U(1)$ gauge field of the D4-brane, which we will set to zero. The D4-brane tension is given by:
\begin{equation}\label{tension}
T_4=\frac{\mu_4}{g_s}=\frac{1}{(2\pi)^4\,\alpha'^{5/2}\,g_s}\ .
\end{equation}
The D4-brane embedding that we consider extends along the radial and
time directions and wraps a three sphere, $S^3$, in the directions
transverse to the D0-brane. To parametrise it let us split the unit
$S^8$ in the metric (\ref{metric}) into:
\begin{equation}
d\Omega_8^2=d\theta^2+\cos^2\theta\,d\Omega_3^2+\sin^2\theta\,d\Omega_4^2\ .
\end{equation}

Our embedding now extends along $t$ and $\Omega_3$ and has a non-trivial profile in the $(u,\theta)$ plane, which we parametrise as $(u,\theta(u))$. Next we Wick rotate the action (\ref{DBI}) and periodically identify time with period $\beta =1/T$. Using equation (\ref{metric}) we obtain:
\begin{equation}\label{DBI-Wick}
S_{\rm DBI}^E =\frac{N_f\,\beta}{8\,\pi^2\,\alpha'^{5/2}\,g_s}\int\, du\,u^3\cos^3\theta(u)\,\sqrt{1+u^2\,f(u)\,\theta'(u)^2}\ .
\end{equation}
In the limit of zero temperature ($u_0\to0$) the regular solution to the equation of motion for $\theta(u)$ is given by $u\,\sin\theta = m$, where the constant $m$ is proportional to the bare mass of the flavours \cite{Karch:2002sh}, \cite{Mateos:2007vn}. At finite temperature the separation $L(u)=u\,\sin\theta(u)$ has a non-trivial profile reflecting the non-vanishing condensate of the theory. To analyse this case it is convenient to define dimensionless radial coordinate $\tilde u = u/u_0$. At large $\tilde u$ the general solution $\theta(\tilde u)$ has the expansion: 
\begin{equation}\label{Expansion-sin(theta)}
\sin\theta =\frac{\tilde m}{\tilde u}+\frac{\tilde c}{u^3}+\dots.
\end{equation}
Holography relates the dimensionless constants $\tilde m, \tilde c$ to the bare mass and condensate of the theory via \cite{Mateos:2007vn}\footnote{Note that our expressions differ slightly from the ones presented in \cite{Mateos:2007vn} due to the different choice of radial variable.}:
\begin{eqnarray}
m_q &=& \frac{u_0\,\tilde m}{2\pi\alpha'}=\left(\frac{120\,\pi^2}{49}\right)^{1/5}\left(\frac{T}{\lambda^{1/3}}\right)^{2/5}\lambda^{1/3}\,\tilde m\ , \nonumber \\
\langle {\cal O}_m\rangle &=&-\frac{N_f\,u_0^3}{2\,\pi\,g_s\,\alpha^{3/2}} \,\tilde c=\left(\frac{2^4 \,15^3\,\pi^6}{7^6}\right)^{1/5}\,N_f\,N_c\,\left(\frac{T}{\lambda^{1/3}}\right)^{6/5}\,(-2\,\tilde c)\ .
\label{dictionary}
\end{eqnarray}
Note that equation (\ref{Expansion-sin(theta)}) implies that the D7-branes are described by a one parameter family of embeddings (parametrised by $\tilde m$). In the case of the D3/D7 system this is natural due to the scaling symmetry (everything depends on the dimensionless ratio $m_q/T$), but is this consistent with the D0/D4 system, which has a dimensionful 'tHooft coupling? Indeed, the D0/D4 system has a 'tHooft coupling $\lambda$ of dimension three suggesting that there are two independent dimensionless parameters $m_q/\lambda^{1/3}$ and $T/\lambda^{1/3}$.  However, while the holographic set-up allows rescaling of the radial coordinate by $u_0$ and the description of D7-brane embeddings by the single parameter $\tilde m$, as can be seen from the first equation (\ref{dictionary}) we have $\tilde m \sim (m_q/\lambda^{1/3}) (T/\lambda^{1/3})^{-2/5}$. As a result the condensate in the second equation in (\ref{dictionary}) is indeed a function of the two dimensionless parameters $(T/\lambda^{1/3},\, m_q/\lambda^{1/3})$ which is consistent with dimensional analysis and is in contrast to the D3/D7 system, where the condensate depends on the dimensionless parameter $m_q/T$.   

\section{Testing the correspondence}
\label{section test}
\begin{figure}[t] 
   \centering
      \includegraphics[width=3.in]{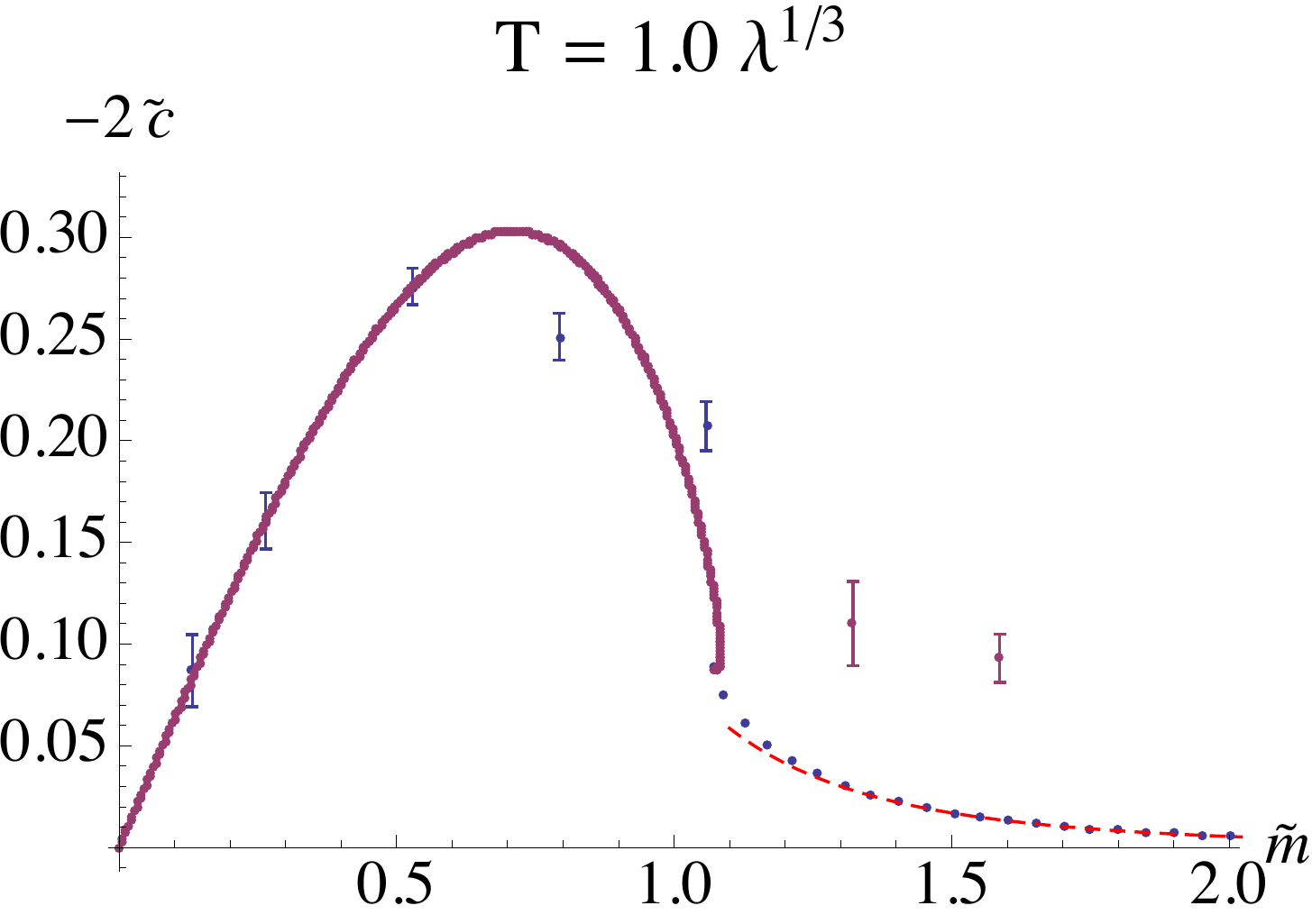} 
   \includegraphics[width=3.in]{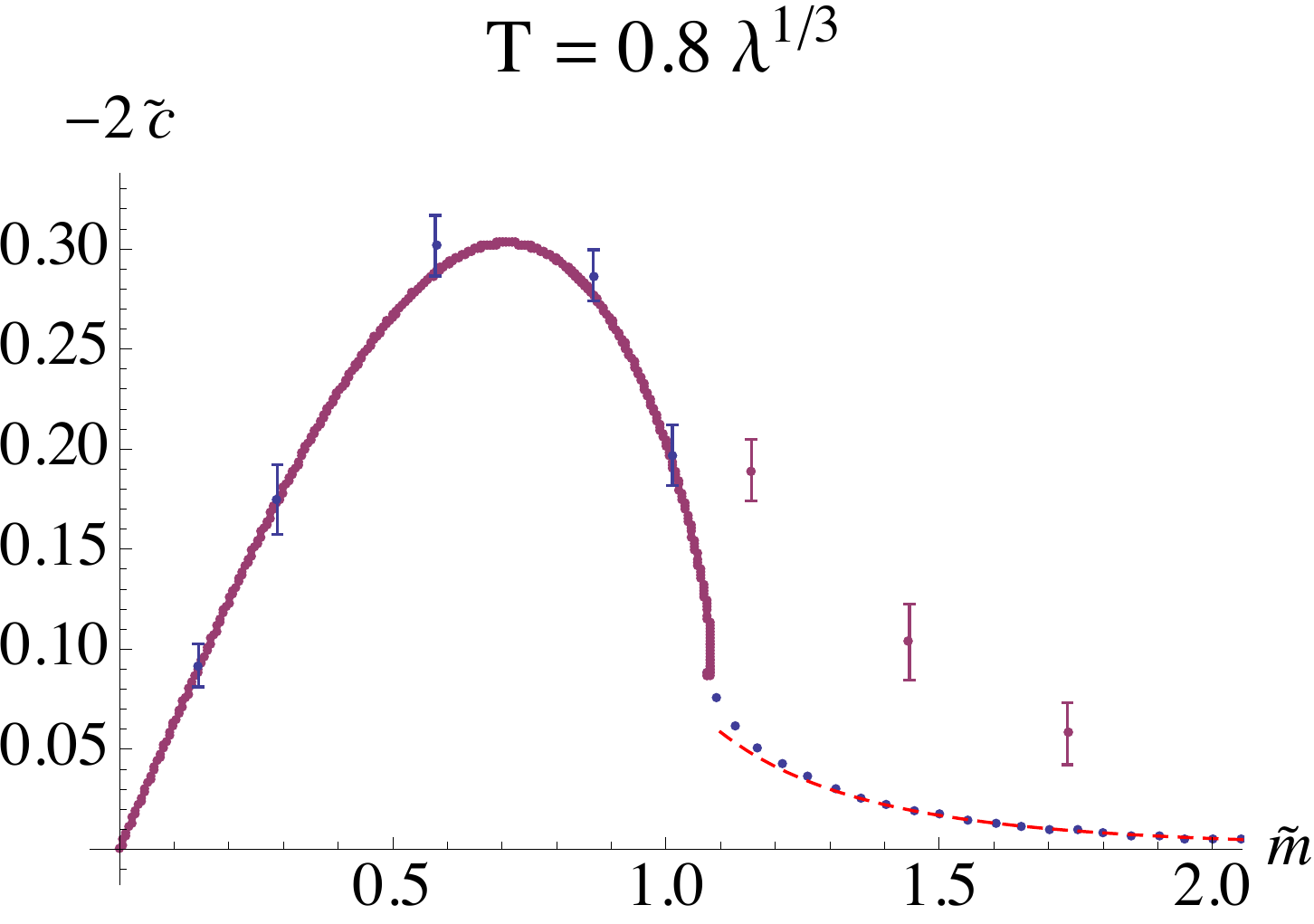} 
 \caption{\small Plots of the condensate versus bare mass parameter curve for $N=10$, $\Lambda =16$ and two different temperatures . {\it left:} For temperature $T = 1.0\,\lambda^{1/3}$ the curve shows excellent agreement at small masses, but deviates quickly from the theoretical curve at greater masses. {\it right:} The curve at temperature $T = 0.8\,\lambda^{1/3}$ exhibits excellent agreement throughout the whole range of masses corresponding to the black hole phase (blue error bars). Similarly to the higher temperature curve there is a significant deviation from the theoretical curve for the Minkowski phase (red error bars).} 
   \label{fig:3}
\end{figure}

In this section we compare the result of the lattice simulations of the model to the predictions of gauge gravity duality. Our main focus is the fundamental condensate of the theory. As definition of the condensate we use the derivative of the free energy of the theory with respect to the bare mass parameter $m_q$.    Note that on the lattice we use a dimensionless mass parameter $m$ and temperature $\tilde T$ given by:
\begin{equation}
m = \frac{m_q}{\lambda^{1/3}}\ ,~~~~\tilde T = \frac{T}{\lambda^{1/3}}\ ,
\end{equation}
where $\lambda$ is the dimensionful 'tHooft coupling. Note also that in $1+0$ dimensions the fundamental condensate $\langle {\cal O}_m\rangle$ is dimensionless. The AdS/CFT dictionary equations (\ref{dictionary}) can be easily rewritten in terms of the lattice dimensionless quantities. One gets:
\begin{eqnarray}
m &=& \left(\frac{120\,\pi^2}{49}\right)^{1/5}\tilde{T}^{2/5}\,\tilde m\ , \nonumber \\
\langle {\cal O}_m\rangle &=&\left(\frac{2^4 \,15^3\,\pi^6}{7^6}\right)^{1/5}\,N_f\,N_c\,\tilde T^{6/5}\,(-2\,\tilde c)\ .
\label{dictionary-lattice}
\end{eqnarray}
 Equations (\ref{dictionary-lattice}) can now be used to scale the dependence $\langle {\cal O}_m\rangle$ versus $m$ obtained from computer simulations and compare to the dependence $-2\,\tilde c$ versus $\tilde m$ obtained from holography. The resulting plots for two temperatures are presented in figure \ref{fig:3}. The plots are for matrix size $N=10$ and lattice spacing $\Lambda = 16$.

The left plot corresponds to temperature $T = 1.0\,\lambda^{1/3}$. One
can observe excellent agreement between the gauge gravity duality and
lattice simulations at small masses ($\tilde m < 1$). However, for
greater masses there is a significant deviation from the theoretical
curve. The right plot corresponds to temperature $T =
0.8\,\lambda^{1/3}$. The excellent agreement between gauge/gravity
predictions and lattice simulations extends for the whole range of
masses within the deconfined (black hole) phase (blue error bars). In
the deconfined (Minkowski) phase there are still significant
deviations from the theoretical curve. These results, we argue
\cite{Filev:2015cmz},
indicate that the $\alpha'$ corrections to the supergravity background
affect the black hole and Minkowski D4-brane embeddings
differently. All black hole embeddings reach the horizon and as a
result experience similar curvature effects for different values of
the mass parameter therefore, the $\alpha'$ corrections largely cancel
when one takes a derivative with respect to the mass to calculate the
condensate. In contrast, Minkowski embeddings close at different
radial distances above the horizon depending on the mass parameter. As
a result the effect of the $\alpha'$ corrections depends strongly on
the mass and contributes to the calculation of the condensate. The
overall better agreement of the lower temperature curve to the
theoretical predictions is another signature that the observed
deviations at large masses are due to $\alpha'$ corrections as opposed
to lattice effects, although at sufficiently high masses ($|m^a|
\lesssim 1/a$) lattice effect also become significant.

\section{Conclusion}

In this paper we review the precision test of holographic flavour
dynamics discussed in detail in \cite{Filev:2015cmz}. We focused on
the study of a one-dimensional flavoured Yang-Mills theory
holographically dual to the D0/D4--brane intersection, also known as
the Berkooz--Douglas matrix model.  We considered a lattice
discretisation of the model and the super-renormalizability of the
model ensures that in the continuum limit, supersymmetry is broken
only by the effect of finite temperature, which enabled us to simulate
it on a computer.

Our results for the condensate versus bare mass curve (which is
universal for different temperatures) show an excellent agreement with
holography in the regime of small bare masses and at lower temperature
this agreement extends to the whole range of masses in the deconfined
phase.  We believe that this agreement can be explained by a
cancellation of the $\alpha'$ corrections to the condensate for black
hole embeddings (deconfined phase). This allows a direct comparison
between computer simulations and AdS/CFT predictions at relatively
high temperatures compared to similar studies of the pure BFSS matrix
model. Furthermore, this remarkable agreement (in the black hole
phase) is obtained without any parameter fitting in contrast to the
analogous studies of the BFSS matrix model~\cite{Hanada:2008ez}, where
the authors performed a fit to estimate the $\alpha'$ corrections to
the internal energy. We believe that it is the cancellation mechanism
described in section \ref{section test}, which allows this highly
non-trivial test of the gauge/gravity correspondence. The improved
agreement as the temperature is reduced is also in accord with this
interpretation.

{\bf Acknowledgements:} We wish to thank Yuhma Asano for useful
comments and discussions. Part of the simulations were performed
within the ICHEC ``Discovery" project ``dsphy003c". The support from
Action MP1405 QSPACE of the COST foundation is gratefully acknowledged.
Veselin Filev is supported in part by the Bulgarian NSF grant DN08/3.

\end{document}